\RequirePackage{fixltx2e}
\documentclass[aip,jcp,reprint,floatfix]{revtex4-1}

\usepackage{amsmath,amsfonts} %
\usepackage{wasysym} %
\usepackage[acronym]{glossaries} %
\usepackage{graphicx} %
\usepackage{dcolumn} %
\usepackage{bm} %
\usepackage{ifpdf} %
\usepackage{color} %
\usepackage{nicefrac} %
\usepackage{natbib} %
\usepackage[byname]{smartref} %
\usepackage[breaklinks, %
colorlinks, %
linkcolor=blue, %
citecolor=blue, %
urlcolor=blue]{hyperref} %
\usepackage[table]{xcolor} %
\usepackage{braket} %

\newcommand{\eq}[1]{Eq.~(\ref{#1})} %
\newcommand{\eqs}[1]{Eqs.~(\ref{#1})} %
\newcommand{\fig}[1]{Fig.~\ref{#1}} %
\newcommand{\bea}{\begin{eqnarray}}
\newcommand{\eea}{\end{eqnarray}}

\begin{document}
                            
\newacronym{CI}{CI}{conical intersection} %
\newacronym{GP}{GP}{geometric phase} %
\newacronym{LVC}{LVC}{linear vibronic coupling} %
\newacronym{DOF}{DOF}{degrees of freedom} %
\newacronym{PES}{PES}{potential energy surface} %
\newacronym{DBOC}{DBOC}{diagonal Born--Oppenheimer correction} %
\newacronym{BMA}{BMA}{bis(methylene) adamantyl} %
\newacronym{FC}{FC}{Franck-Condon} %
\newacronym{CWE}{CWE}{cylindrical wave expansion}
\newacronym{DVBC}{DVBC}{double value boundary conditions}

\title{A diabatic definition of geometric phase effects}

\author{Artur F. Izmaylov} %
\affiliation{Department of Physical and Environmental Sciences,
  University of Toronto Scarborough, Toronto, Ontario, M1C 1A4,
  Canada} %
\affiliation{Chemical Physics Theory Group, Department of Chemistry,
  University of Toronto, Toronto, Ontario M5S 3H6, Canada} %
\author{Jiaru Li}
\affiliation{Department of Physical and Environmental Sciences,
  University of Toronto Scarborough, Toronto, Ontario, M1C 1A4,
  Canada} %
\author{Lo{\"i}c Joubert-Doriol} %
\affiliation{Department of Physical and Environmental Sciences,
  University of Toronto Scarborough, Toronto, Ontario, M1C 1A4,
  Canada} %
\affiliation{Chemical Physics Theory Group, Department of Chemistry,
  University of Toronto, Toronto, Ontario M5S 3H6, Canada} %

\date{\today}

\begin{abstract}
Electronic wave-functions in the adiabatic representation acquire nontrivial geometric phases (GPs) 
when corresponding potential energy surfaces undergo conical intersection (CI). These GPs 
have profound effects on the nuclear quantum dynamics and cannot be eliminated in the 
adiabatic representation without changing the physics of the system. 
To define dynamical effects arising from the GP presence the nuclear quantum dynamics 
of the CI containing system is compared with that of the system with artificially 
removed GP. We explore a new construction of the system with removed GP via a modification of 
the diabatic representation for the original CI containing system. 
Using an absolute value function of diabatic couplings we remove the GP while preserving 
adiabatic potential energy surfaces and CI. 
We assess GP effects in dynamics of a two-dimensional linear vibronic coupling model 
both for ground and excited state dynamics. Results are compared with those 
obtained with a conventional removal of the GP by ignoring double-valued boundary 
conditions of the real electronic wave-functions. Interestingly, GP effects appear similar in 
two approaches only for the low energy dynamics. In contrast with the conventional approach, 
a new approach does not have substantial GP effects in the ultra-fast 
excited state dynamics.    
\end{abstract}


\maketitle

\glsresetall

%%%%%%%%%%%%%%%%%%%%%%%%%%%%%%%%%%%%%%%%%%%%%%
%\sloppy
\section{Introduction}
\label{sec:introduction}

%$\nabla_{\mathbf{R}}$
%1. When do we have GP - CIs, adiabatic representation 

Ubiquitous in molecules beyond diatomics, \glspl{CI} of electronic states act as 
``funnels''~\cite{Balzer:2003/cpl/351,Koppel:1984/acp/59, Yarkony:1996/rmp/985,
  Hahn:2000/jpcb/1146} that enable rapid
conversion of the excessive electronic energy into nuclear motion. 
%Owing to the ubiquity of \glspl{CI} in molecules,~\cite{Cederbaum:1977/cp/169,
%  Seidner:1993/cpl/117,Woywod:1994/jcp/1400, Kendrick:1996/jcp/7502,
%  Hahn:2000/jpcb/1146, Cattarius:2001/jcp/2088,
 % Vallet:2005/jcp/144307, Burghardt:2008/jcp/174104,
 % Sardar:2008/pccp/6388, Sirjoosingh:2011/jctc/2831,
 % Domcke:2012/arpc/325, Halasz:2012/jpca/2636, Ou:2013/jpcc/19839}
% an adequate theoretical description of this conversion mechanism is an
% important task in theoretical physical chemistry.
Also, CIs lead to the appearance of the \gls{GP}~\cite{Berry:1984/rspa/45,
Mead:1979/jcp/2284, Berry:1987/rspa/31} in both electronic and nuclear wave-functions
of the adiabatic representation. 
The  \gls{GP} presence leads to a sign change of adiabatic electronic wave-functions 
along a closed path of nuclear configurations encircling the \gls{CI} 
seam.\cite{LonguetHigg:1958/rspa/1,Mead:1979/jcp/2284}
This sign change affects evaluation of nonadiabatic couplings (NACs) necessary to 
complete the nuclear kinetic energy part of the adiabatic representation to define
a nuclear Schr\"odinger equation. Changes in NACs due to the GP can lead 
to profound modification of nuclear dynamics even in situations when the nuclear 
wave-function is localized far from the region of CI. For example, 
the \gls{GP} causes an extra phase accumulation for fragments of the nuclear 
wave-packet that move around the \gls{CI} on opposite 
sides.\cite{Schon:1995/jcp/9292,Ryabinkin:2013/prl/220406} 
This leads to destructive interference that gives rise either to a spontaneous localization of the nuclear
density~\cite{Ryabinkin:2013/prl/220406} or slower nuclear dynamics
\cite{Loic:2013/jcp/234103} than in the case where the \gls{GP}
is neglected. 

To distinguish unambiguously what is the effect of the GP on the 
nuclear dynamics one can study the exact
quantum dynamics, which necessarily incorporates all GP effects, in comparison with the 
dynamics that is not including the GP. This comparison would allow one to formulate unique 
dynamical features related to the CI topology which gives rise to the GP.
A natural question is how to modify a computational 
scheme to remove the GP with a minimal effect on other parts of dynamics?
Previously, to analyze GP effects %\cite{refs on people who did this} 
constructing a GP excluded version has been done by switching 
to the adiabatic representation.\cite{Hazra:2015he,Kendrick:1996/jcp/7475,Kendrick:2003,Althorpe:2008/jcp/214117,Baer:2000/cp/123} A straightforward simulation of the nuclear 
dynamics ignoring double-valued character of electronic and nuclear 
wave-functions in the adiabatic representation excludes the GP.\cite{Mead:1979/jcp/2284} 
As shown by Mead and Truhlar, 
the only change that is needed to obtain the correct nuclear dynamics 
in the adiabatic representation is a phase modification for both electronic and nuclear 
wave-functions that returns single-valued boundary conditions to these functions.\cite{Mead:1979/jcp/2284} 
This phase change modifies only the kinetic energy terms, NACs, in the nuclear Hamiltonian 
and leaves potential energy terms unchanged.
A practical difficulty with this approach is that it requires performing quantum nuclear dynamics 
in the adiabatic representation where many NAC components diverge at the CI. 
The necessity to work in the adiabatic representation 
creates technical challenges for investigation of GP effects in realistic systems beyond 
low dimensional simple models.

In this paper we propose an alternative way of investigating GP effects by introducing 
a modification in the system diabatic Hamiltonian, this modification removes the GP in the corresponding 
adiabatic representation without altering potential energy surfaces. Our modification is not equivalent 
to ignoring double-valued boundary conditions in the adiabatic representation and provides 
a new set of results characterizing GP effects in CI problems. 

The rest of the paper is organized as follows. In Sec. \ref{sec:theoretical-analysis} 
we introduce our approach for a two-dimensional linear vibronic coupling 
model problem with CI. Section \ref{sec:numerical-examples} provides numerical results 
comparing GP effects obtained in the new diabatic and old adiabatic approaches on a 
set of model systems  parametrized using real molecular systems. Finally, Sec. \ref{sec:concl}  
concludes the work by summarizing main results. 

%%%%%%%%%%%%%%%%%%%%%%%%%%%%%%%%%%%%%%%%%%%%%%%%%%%%%%%%%%%%%%%%%%%%%%
\section{Theoretical analysis}
\label{sec:theoretical-analysis}

We introduce two models within the two-dimensional \gls{LVC} Hamiltonian
\begin{equation}
  \label{eq:H_lvc}
  \hat H_{\rm LVC} = {\hat T}  {\mathbf 1}_2 + 
  \begin{pmatrix} 
    V_{11} & V_{12} \\
    V_{12} & V_{22}
  \end{pmatrix},
\end{equation}
where $\hat T = -\frac{1}{2}\nabla^2 \equiv -\frac{1}{2}(\partial^2
/\partial x^2 +\partial^2 /\partial y^2) $ is the nuclear kinetic
energy operator, and ${\mathbf 1}_2$ is a $2\times 2$ unit matrix.
\footnote{Atomic units will be used throughout this paper.} 
 $ V_{11}$ and
$ V_{22}$ are the diabatic potentials represented by identical 2D
parabolas shifted in the $x$-direction by $a$ and in energy by $\Delta$
\begin{align}
  \label{eq:diab-me-11}
  V_{11} = {} & \frac{\omega_1^2}{2}x^2
  + \frac{\omega_2^2}{2}y^2 ,\\
  \label{eq:diab-me-22}
  V_{22} = {} & \frac{\omega_1^2}{2}\left(x - a\right)^2 +
  \frac{\omega_2^2}{2}y^2 -\Delta.
\end{align}
To have the \gls{CI} in the adiabatic representation, $V_{11}$ and
$V_{22}$ are coupled by a linear potential $V_{12}=c y$ in model 1
 and by an absolute value of a linear potential $V_{12}=c|y|$ in model 2.

\begin{figure}
  \centering
  \includegraphics[width=0.5\textwidth]{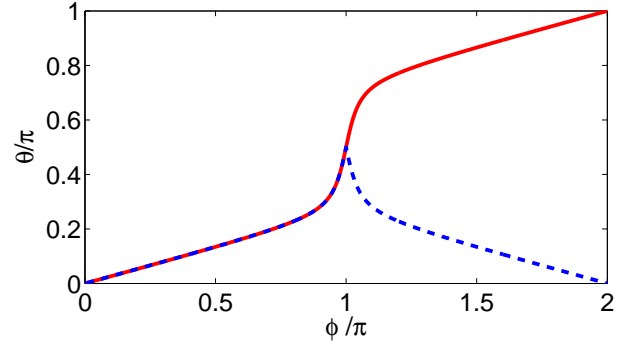}
  \caption{$\theta$ angle of the diabatic-to-adiabatic transformation 
  as a function of the CI encircling angle $\phi$ for two models: red solid for model 1 [\eq{eq:theta1}] 
  and blue dashed for model 2 [\eq{eq:theta2}].}
  \label{fig:circ}
\end{figure}

%Hamiltonian~\eqref{eq:H_lvc} is written in mass-weighted effective
%coordinates for a subsystem that exhibits a \gls{CI}. 
%\subsection{Adiabatic representation of the 2D LVC model}
%\label{sec:adiab-repr-2d}
Switching to the adiabatic representation for the 2D LVC Hamiltonian in 
\eq{eq:H_lvc} is done by diagonalizing the potential matrix 
using a unitary transformation 
\begin{equation}
  U = 
  \label{eq:Umat}
  \begin{pmatrix}
    \cos\theta & \sin\theta \\
    -\sin\theta & \cos\theta
  \end{pmatrix}
\end{equation}
that introduces adiabatic electronic states
\begin{eqnarray}\label{eq:adi1}
  \ket{\phi_1^\text{adi}} & = & \phantom{-}\cos\theta\,\ket{\phi_1} +
  \sin\theta\,\ket{\phi_2}, \\ \label{eq:adi2}
  \ket{\phi_2^\text{adi}} & = & -\sin\theta\,\ket{\phi_1} +
  \cos\theta\,\ket{\phi_2},
\end{eqnarray}
with $\theta=\theta(x,y)$ as a rotation angle between the diabatic electronic
states $\ket{1}$ and $\ket{2}$
\begin{equation}
  \label{eq:theta}
  \theta = \frac{1}{2}\arctan \dfrac{2\,V_{12}}{V_{22} - V_{11}}. %=
  %\frac{1}{2}\arctan \dfrac{\gamma y}{x + b}.
\end{equation}
The transformation in \eq{eq:Umat} gives rise to the 2D \gls{LVC}
Hamiltonian in the adiabatic representation $\hat H_\text{adi}  = U^\dagger
{\hat H}_{\rm LVC} U$,
\begin{equation}
  \label{eq:adiab}
   \hat H_\text{adi}  =   
  \begin{pmatrix}
    \hat T + \hat\tau_{11}& \hat\tau_{12} \\
    \hat\tau_{21} & \hat T +\hat\tau_{22}
  \end{pmatrix} +
  \begin{pmatrix}
    W_{-} & 0 \\
    0 & W_{+}
  \end{pmatrix},
\end{equation}
where
\begin{align}
  \label{eq:Wmin}
  W_{\pm} = & {} \dfrac{1}{2}\left(V_{11} + V_{22}\right) \pm
  \dfrac{1}{2}\sqrt{\left(V_{11} - V_{22}\right)^2 + 4 V_{12}^2}
\end{align}
are the adiabatic potentials which are exactly the same for models 1 and 2,
and 
$\hat\tau_{ij} = -\bra{\phi_i^\text{adi}}\nabla\phi_j^\text{adi}\rangle\nabla-\bra{\phi_i^\text{adi}}\nabla^2\phi_j^\text{adi}\rangle/2$ are the nonadiabatic
couplings. For two-electronic-state models we can express $\hat\tau_{ij}$ as
\begin{align}
  \label{eq:tau-adi-diag}
  \hat\tau_{11} & {} =\hat\tau_{22} = \frac{1}{2}
  \nabla\theta\cdot\nabla\theta  \\ \label{eq:tau-adi-offd}
  \hat\tau_{12} & {} = -\hat\tau_{21} =
  \frac{1}{2}\left(\nabla^2\theta+2\nabla\theta\cdot\nabla\right).
\end{align}
The diagonal non-adiabatic couplings, $\hat\tau_{11}$ and $\hat\tau_{22}$, represent a repulsive
potential known as the \gls{DBOC}.\cite{Born:1954,Handy:1996/cpl/425,
  Valeev:2003/jcp/3921}  The off-diagonal elements, $\hat\tau_{12}$ and
$\hat\tau_{21}$ in Eq.~\eqref{eq:tau-adi-offd}, couple dynamics on 
the adiabatic potentials $W_{\pm}$ and are responsible for 
non-adiabatic transitions. All $\hat\tau_{ij}$ terms involve derivative of 
$\theta$ which is given by two different functions 
\bea
  \label{eq:theta1}
  \theta_1 &=& \frac{1}{2}\arctan \dfrac{\gamma y}{x - b}, \\  \label{eq:theta2}
  \theta_2 &=& \frac{1}{2}\arctan \dfrac{\gamma |y|}{x - b}
\eea
for models 1 and 2, respectively.
Here, $b = \Delta/(\omega_1^2 a)$ is the $x$-coordinate of the CI
point, and $\gamma = {2c}/{(\omega_1^2a)}$ is dimensionless coupling
strength. For simplicity of the subsequent analysis we set $b =0$,
which corresponds to centring the coordinates at the \gls{CI} point.
To see the difference between $\theta_1$ and $\theta_2$ 
we will continuously track their changes along a contour encircling the CI. %(\fig{fig:circ}). 
For the CI located at the origin we have taken a set of points on a circle $(x_j,y_j)$ 
parametrized by the polar representation of complex numbers $x_j+iy_j = re^{i\phi_j}$, where 
$r=1$ and $\phi_j$'s are taken from the discretized $[0,2\pi]$ interval. 
Figure~\ref{fig:circ} illustrates that $\theta_1$ changes by $\pi$ when we do the 
full circle while $\theta_2$ returns to its initial value, 0.  
 For the adiabatic electronic functions [\eqs{eq:adi1}-(\ref{eq:adi2})] 
 this means that these functions change their signs in model 1 
 and return to their original values in model 2. Therefore, models 1 and 2 have 
 electronic functions which are double- and single-valued functions of nuclear parameters, 
 respectively. In terms of differentiability, $\theta_2$ clearly has issues at the $y=0$ line. However,
we will not compute $\hat \tau_{ij}$ elements for model 2 because all simulations for this model 
will be done in the diabatic representation. 

Another possible concern for our approach could be that the modification of the diabatic model removing 
the GP breaks smoothness of the diabatic coupling as a function of the nuclear coordinate. 
This raises a question of the physical meaning of the diabatic model with such a coupling term. 
It is important to understand that the GP is a significant part of the CI topology and removing it in any 
way is expected to produce an incomplete and thus in some sense unphysical picture. 
To illustrate this point even further 
we will show that the diabatic model which is mathematically equivalent to the adiabatic model with 
the GP removed in the conventional way has divergent diabatic potentials with discontinuous derivatives. 
%that corresponds to the conventional 
%adiabatic Hamiltonian in the space of single-valued function and thus ignoring the GP.
 First, let us clarify that to obtain the adiabatic Hamiltonian that will produce results equivalent to the initial diabatic 
LVC Hamiltonian [\eq{eq:H_lvc}] in the space of single-valued functions one needs to use the following 
single-valued transformation $U^{(1)} = e^{i\theta} U$.
Note that both functions $e^{i\theta}$ and $U$ [\eq{eq:Umat}] in this product are double-valued 
but they give the single-valued resulting transformation. 
In contrast to $U$, $U^{(1)}$ allows us to move between the representations while staying in the space of single-valued functions,
hence, $\hat{H}_{\rm adi}^{(1)} = {U^{(1)}}^{\dagger} \hat H_{\rm LVC}U^{(1)}$ is the proper adiabatic Hamiltonian 
in the space of the single-valued functions. 
To generate the diabatic counterpart of the conventional Hamiltonian $\hat H_{\rm adi}$ [\eq{eq:adiab}] one 
should also use $U^{(1)}$ but for the inverse transformation
\bea
\hat H_{\rm dia} &=& U^{(1)} \hat H_{\rm adi}{U^{(1)}}^{\dagger} \\ \label{eq:Hdia}
&=& \hat H_{\rm LVC} +  \left(\frac{1}{2}(\nabla\theta)^2 +\frac{i}{2}\nabla^2\theta+i\nabla\theta\nabla \right) \mathbf{1}_2.
\eea
$\hat H_{\rm dia}$ is similar to $\hat H_{\rm LVC}$ but it has an extra term containing derivatives of the mixing angle $\theta$.
It is well known that all these derivatives diverge at the CI point\cite{Ryabinkin:2014/jcp/214116} 
thus giving rise to the diabatic representation 
that is unphysical. For example, there are two potential-like terms 
in \eq{eq:Hdia}, $\frac{1}{2}(\nabla\theta)^2+\frac{i}{2}\nabla^2\theta$, 
which can be formally considered as a modification of diabatic surfaces $V_{11}$ and $V_{22}$.
This modification produces divergent diabatic surfaces with nuclear derivative discontinuities.  
All these problems in the diabatic representation of the conventional  way of the GP removal 
has not been discussed before because the diabatic Hamiltonian $\hat H_{\rm dia}$ 
does not provide any advantage compare to its adiabatic counterpart $\hat H_{\rm adi}$
and thus has not been used in simulations. 
This example illustrates that although introducing the absolute value of the coupling term 
leads to nuclear derivative discontinuities, this modification is still better than the conventional approach 
with its divergent diabatic potential terms.

%%%%%%%%%%%%%%%%%%%%%%%%%%%%%%%%%%%%%%%%%%%%%%%%%%%%%%%
\section{Numerical examples}
\label{sec:numerical-examples}

We will consider three molecular systems with \glspl{CI}
 that are well described by multi-dimentional \gls{LVC} models: the
\gls{BMA}~\cite{Izmaylov:2011/jcp/234106}
and butatriene~\cite{Koppel:1984/acp/59,Ryabinkin:2014/jcp/214116} cations, and the
pyrazine molecule.~\cite{Burghardt:2008/jcp/174104,Ryabinkin:2014/jcp/214116} 
$N$-dimensional \gls{LVC} models for these systems are taken 
from literature\cite{Izmaylov:2011/jcp/234106,Cattarius:2001/jcp/2088,Raab:1999/jcp/936}. 
Although our approach to removing the GP can be easily applied to a 
multi-dimensional LVC, for the sake of simplicity and also to be able to compare with 
our previous simulations\cite{Ryabinkin:2014/jcp/214116} 
we will use 2D effective LVC Hamiltonians 
for these systems (see Table~\ref{tab:BMA-param}).
\begin{table}
  \caption{Parameters of the 2D effective \protect\gls{LVC}
    Hamiltonian, Eq.~\eqref{eq:H_lvc}, for the studied systems, 
    and the $x$-coordinate of the Franck-Condon point ($x_{\rm FC}$). 
    The $y$-coordinate of the Franck-Condon point is zero. } 
  \label{tab:BMA-param}
  \centering
  \begin{ruledtabular}
    \begin{tabular}{@{}lccccr@{}}
      \multicolumn{1}{c}{$\omega_1$} & $\omega_2$ & $a$ & $c$ &
      \multicolumn{1}{c}{$\Delta$} & $x_{\rm FC}$\\ \hline
      \multicolumn{6}{c}{ Bis(methylene) adamantyl cation} \\
      $7.743\times10^{-3}$ & $6.680\times10^{-3}$ & 31.05 &
      $8.092\times 10^{-5}$ & 0.000 & 0.000 \\[1ex]
      \multicolumn{6}{c}{ Butatriene cation} \\
      $9.557\times10^{-3}$ & $3.3515\times10^{-3}$  & 20.07   &
      $6.127\times 10^{-4}$ & 0.020  & 6.464\\[1ex]
      \multicolumn{6}{c}{ Pyrazine} \\
      $3.650\times10^{-3}$ & $4.186\times10^{-3}$ & 48.45 & $4.946\times
      10^{-4}$ & 0.028 & 29.684
    \end{tabular}
  \end{ruledtabular}
\end{table}
To quantify \gls{GP} effects we solve the time-dependent nuclear 
Schr\"odinger equation for three model Hamiltonians: 
1) model 1 using the diabatic representation (Diab-wGP) 2) model 2 using the diabatic 
representation (Diab-noGP), and 3) model 1 using the adiabatic representation [\eq{eq:adiab}] 
and ignoring double valued character of electronic and nuclear wave-functions (Adiab-noGP).  
First two Hamiltonians were treated using the split-operator approach while for 
the third one the exact diagonalization in a finite basis was employed.\cite{Ryabinkin:2014/jcp/214116}
In what follows we will consider two dynamical regimes different in energy of an initial wave-packet: 
1) low energy case, where dynamics mostly occurs near CI on the ground electronic state; 
2) high energy case, when a wave-packet proceeds from the excited electronic state to the ground 
state through the CI.
%Two sets of initial conditions were investigated, placing a wave-packet either on the ground or on the 
%excited electronic state.

\subsection{Low energy dynamics}
\label{sec:gsdyn}

\begin{figure}
  \centering
  \includegraphics[width=0.5\textwidth]{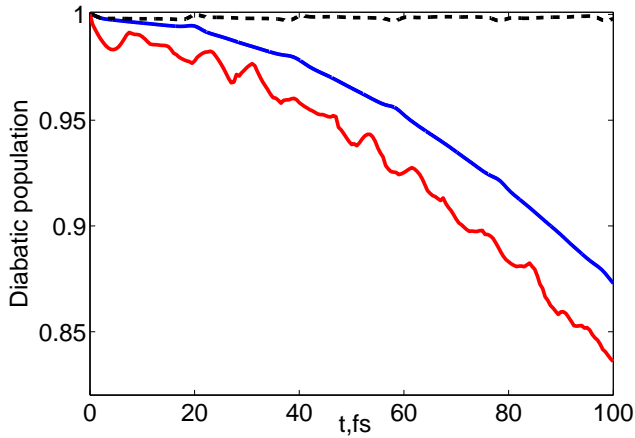}
  \caption{Diabatic population dynamics of the \protect\gls{BMA} cation: 
 Diab-wGP (dashed black), Adiab-noGP (solid red), Diab-noGP (solid blue).}
  \label{fig:bma}
\end{figure}

For low energy dynamics we will analyze only the 
BMA case because the other systems have a non-symmetric diabatic well structure
that would freeze dynamics if one starts in the lower energy well.
The ground vibrational state of the uncoupled $V_{11}$ diabatic  
potential  
\begin{equation}
  \label{eq:Gaussian_wp}
  \chi(x,y) = \frac{(\omega_1\omega_2)^{1/4}}{\pi^{1/2}}
  \exp{\left(-\frac{\omega_1(x-x_{\rm FC})^2}{2} -
      \frac{\omega_2y^2}{2}\right)}
\end{equation} 
was chosen as an initial wave-packet.
%with widths $\sigma_x =\sqrt{2/\omega_1}$ and $\sigma_y = \sqrt{2/\omega_2}$.
The diabatic population of the initial state is monitored as a function of time to assess 
dynamics (\fig{fig:bma}), this 
population correlates well with the well population in the adiabatic representation for BMA.

For discussing diabatic population evolution (\fig{fig:bma}) it is convenient to introduce a notation 
for diabatic uncoupled vibrational levels, $(n,m)_{s}$ refers to a level with $n$ vibrational
quanta on the $x$ (tuning) coordinate and $m$ vibrational quanta on the 
$y$ (coupling) coordinate for the diabatic state $s=D,A$. 
$s=D(A)$ will correspond to $V_{11}(V_{22})$ diabats.
In this notation the initial state is $(0,0)_D$ and in model 1 it is coupled only with 
$(n,1)_A$ states, where $n$ is any positive integer number. 
Since all $(n,1)_A$ states are higher in energy 
than $(0,0)_D$, the transfer is negligible in the Diab-wGP method. 
On the other hand, in model 2, owing to the even coupling 
function $c|y|$, the initial state $(0,0)_D$ is coupled with $(n,2k)_A$ states, 
where $n$ and $k$ are arbitrary integer numbers. 
Thus there is a resonance channel $(0,0)_D\rightarrow(0,0)_A$ 
that is responsible for a donor population decay quadratic in time 
in the Diab-noGP method. These results can be also obtained using 
the time-dependent perturbation theory which is applicable here due to a small value 
of the coupling constant, $c$. Both Diab-wGP and Diab-noGP methods 
have small bumps on the population plot with the period of 20 fs corresponding to the 
tuning coordinate frequency $\omega_1=2\pi/20$ fs$^{-1}$. 
These features come from off-resonance  
transitions $(0,0)_D\rightarrow(n,1)_A$ and $(0,0)_D\rightarrow(n,2k)_A$ for $n\ge 1$ in 
Diab-wGP and Diab-noGP methods, respectively. Using the time-dependent perturbation
theory and summation over states of harmonic oscillators it can be shown 
that the off-resonance channel should induce the population dynamics with 
a frequency corresponding to $\omega_1$.\cite{Izmaylov:2011/jcp/234106} 
The Adiab-noGP method has very similar dynamics as that in Diab-noGP. This can be attribute to the 
absence of destructive interference between two pathways around the CI  located 
between the wells when we ignore the double-valued boundary conditions by using the Adiab-noGP approach. 
Thus, in Adiab-noGP, one observes coherent tunnelling between 
the wells as in any single electronic state double-well problem.

\subsection{Excited state dynamics}
\label{sec:exdyn}

\begin{figure}
  \centering
  \includegraphics[width=0.5\textwidth]{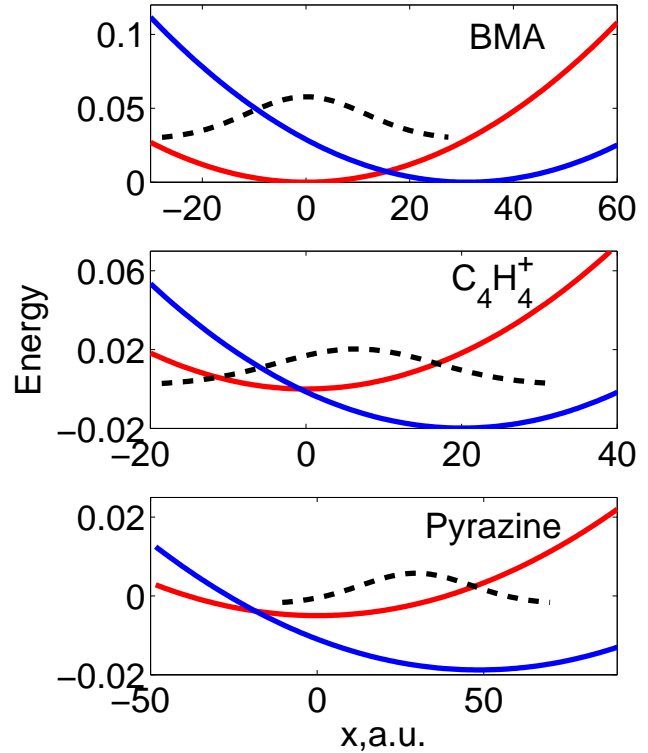}
  \caption{$y=0$ cuts of the diabats\cite{note:DA}(red and blue) and the initial wave-packet (black dashed) for excited
  state dynamics of  \protect\gls{BMA} cation, $\rm C_4H_4^{+}$, 
  and pyrazine.}
  \label{fig:PESx}
\end{figure}

\begin{figure}
  \centering
  \includegraphics[width=0.5\textwidth]{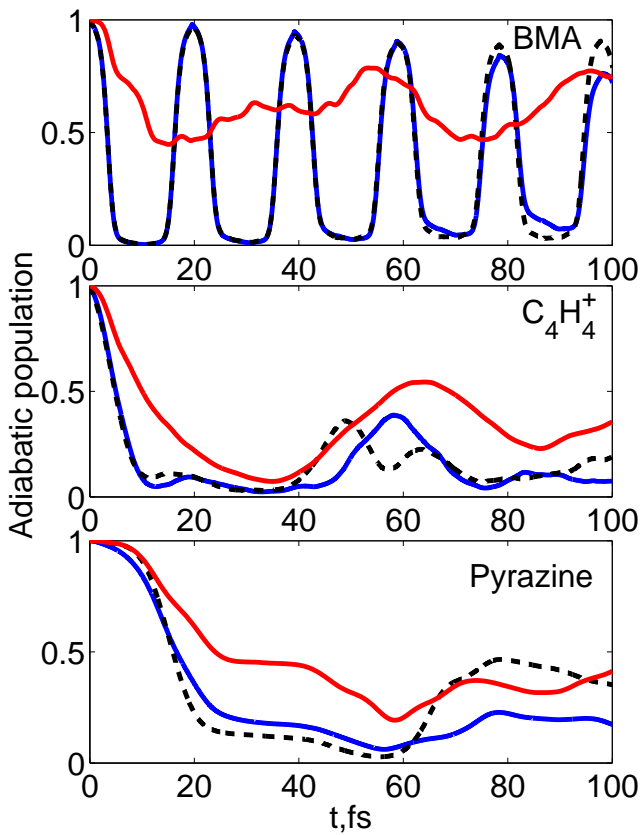}
  \caption{Excited state population dynamics of \protect\gls{BMA} cation, $\rm C_4H_4^{+}$, 
  and pyrazine: Diab-wGP (dashed black), Adiab-noGP (solid red), Diab-noGP (solid blue).}
  \label{fig:ex_dyn}
\end{figure}

All three systems presented in Table I are assessed here so that results of 
our previous study\cite{Ryabinkin:2014/jcp/214116} using the 
Adiab-noGP approach can be contrasted with those of Diab-noGP.
A Gaussian wave-packet [\eq{eq:Gaussian_wp}]
centred at a \gls{FC} point and placed on the excited adiabatic electronic state is taken as 
an initial nuclear wave-function (Table \ref{tab:BMA-param} and \fig{fig:PESx}).
The quantity characterizing excited state  dynamics will be  the adiabatic electronic state population 
$P_{\rm adi}(t)  =  \braket{\chi^\text{adi}_2(t)|\chi^\text{adi}_2(t)}$,
where $\chi^\text{adi}_2(x,y,t)$ is a time-dependent nuclear wave-function
that corresponds to the excited adiabatic electronic state (\fig{fig:ex_dyn}).

%All three figures with 1) exact 2) no GP in adiabatic representation 3) no GP with the absolute value 
%$V_{12}=c|y|$ coupling.

For BMA, due to low diabatic coupling, 
the exact dynamics (Diab-wGP) corresponds to coherent oscillations on a 
donor diabatic surface. Once the wave-packet 
crosses the diabatic state intersection the adiabatic population switches from excited to
the ground state, but the wave-packet resides almost completely on the same diabat. 
The period of these oscillations corresponds exactly to the tuning mode 
frequency $\omega_1=2\pi/20$ fs$^{-1}$. 
Switching to the Diab-noGP approach does not change dynamics within a 
sub 100 fs time-scale because 
small $c$ makes transitions between diabatic levels inefficient. In other words, the difference in 
the coupling structure $(n,m)_s\rightarrow(n',m\pm1)_{s'}$ for model 1 versus 
$(n,m)_s\rightarrow(n',m\pm2k)_{s'}$ for model 2 does not cause large differences in 
population dynamics until population transfer between diabatic states becomes
appreciable. Differences between results of Adiab-noGP and Diab-wGP have 
been extensively discussed in Ref.~\onlinecite{Ryabinkin:2014/jcp/214116},  
and in BMA, they correspond to compensation of DBOC by GP induced terms in NACs.  
Without GP, DBOC has a significant repulsive character that prevents the wave-packet 
from approaching a CI region and thus hinders nonadiabatic transfer. 

In the butatriene cation and pyrazine, the initial wave-packets are 
much closer to the CI (\fig{fig:PESx}) and diabatic coupling constant $c$ is more 
than 5 times larger than in the BMA case. 
Thus, the time-scale of the adiabatic population dynamics is regulated by 
the nonadiabatic transition rather than oscillations
on a diabatic surface. Pyrazine due to its further FC point from the CI has a small plateau 
region in the initial population dynamics, this plateaux corresponds to a wave-packet 
approach to the CI. 
As in the BMA case, differences between Diab-wGP and Diab-noGP appear at a longer time-scale 
than that of the initial nonadiabatic transition. 
Absence of the difference in Diab-wGP and Diab-noGP 
can be attributed to averaging over transitions of many diabatic vibrational states 
forming a wave-packet on the excited state.  These vibrational states 
although individually may have some differences in transferring population to accepting states 
in two models, but for the overall transfer such differences are averaged out. 
%Any vibrational diabatic level $(n,m)_s$ is coupled with all $(n',m\pm2k)_{s'}$, where $n'$ and $k$ are any integer numbers for the absolute coupling and with $(n',m\pm1)_{s'}$ levels for the linear coupling. 
The difference between Adiab-noGP and Diab-wGP is apparent 
even at ultra-fast initial transitions and has origin in 
enhancement of nonadiabatic transfer due to the GP 
for some parts of the nuclear wave-packet.\cite{Ryabinkin:2014/jcp/214116}

%%%%%%%%%%%%%%%%%%%%%%%%%%%%%%%%%%%%%%%%%%%%%%%%%%%%%%%%%%%%%%%%%%%%%%
\section{Concluding remarks}
\label{sec:concl}

We presented a new method of analyzing GP induced effects in dynamics. It has conceptually important 
aspects and practical advantages. Conceptually, it is interesting to see what are the 
possible ways to remove the GP and how different these ways are in terms of quantum dynamics.
Previously, to remove the GP one could ignore double-valued boundary conditions of 
electronic and nuclear wave-functions, this led to modifying both low energy dynamics 
and fast excited state dynamics. The new approach shows the same effect of the GP removal 
for the low energy dynamics, but does not have substantial effect in the fast excited state dynamics.
Practically, the new approach gives an opportunity to study GP effects in the diabatic representation 
where simulation methods are much more developed (e.g., Multi-configuration time-dependent
Hartree approach). Thus we can easily 
explore $N$-dimensional scenarios without necessity for additional transformations. 
Going beyond linear vibronic coupling is also possible because our main 
modification puts absolute value on the coupling term so that in the two-electronic 
state problem $V_{12}$ transforms into $|V_{12}|$ without changing the adiabatic 
potential energy surfaces. 

%Also,
%going to higher powers of coupling coordinates can help to restore differentiability of the 
%mixing angle functions. For example, $y|y|$ vs $y^2$ or $y^3$ vs $|y^3|$ in both examples 
%one function will produce GP and the other won't but all these functions at least one-time 
%differentiable. 

%%%%%%%%%%%%%%%%%%%%%%%%%%%%%%%%%%%%%%%%%%%%%%%%%%%%%%%%%%%%%%%%%%%%%%
\section{Acknowledgments}

A.F.I. acknowledges funding from a Sloan Research Fellowship 
and the Natural Sciences and Engineering Research Council  
of Canada (NSERC) through the Discovery Grants Program.
L.J.D. is grateful to the European Union Seventh Framework Programme
(FP7/2007-2013) for the financial support under grant agreement PIOF-GA-2012-332233.

%\bibliography{gp}
%

\end{document}